\documentstyle[epsf]{lamuphys}
\begin{document}
\title{Diffusive and subdiffusive step dynamics}

\author{W.\,Selke\inst{} \and M.\,Bisani\inst{}}

\institute{Institut f\"ur Theoretische Physik, Technische 
Hochschule, D-52056 Aachen, Germany}

\maketitle

\begin{abstract}
The dynamics of steps on crystal surfaces is considered. In general, the
meandering of the steps obeys a 
subdiffusive behaviour. The characteristic asymptotic
time laws depend on the 
microscopic mechanism for detachment and attachment of the atoms
at the steps. The three limiting cases of
step--edge diffusion, evaporation--condensation
and terrace diffusion are studied in the framework of Langevin
descriptions and by Monte Carlo simulations.
\end{abstract}
\section{Introduction}
Surface dynamics has attracted much interest in recent years, partly
due to progress in experimental techniques to measure
atomic processes at surfaces. A variety of phenomena has been
studied, both experimentally and theoretically, including
fluctuations of isolated and bunches of steps, island motion, and
thermal relaxation of sinusoidal
gratings (\cite{vil:1,pi:1,wi:1,sel:1,ha:sp,k:e2,bon:zel,w:1,k:e1}).
 
In this contribution, we shall draw attention to the 
(sub--)diffusive dynamics of the meandering of 
steps of monoatomic height. We 
shall emphasize the role of different atomic mechanisms leading
to distinct limiting cases for the asymptotic time laws. In particular, (i)
evaporation--condensation (EC), i.e. uncorrelated attachment
and detachment of atoms at the step edge, (ii) periphery
diffusion (PD) along the step edge, and (iii) terrace
diffusion (TD) will be discriminated (these three cases
correspond to evaporation--condensation, surface diffusion and
volume diffusion in the pioneering work of Mullins (1959) on
thermal flattening of crystal surfaces). Results of Langevin descriptions
and of Monte Carlo simulations will be presented.
 
\section{Step Fluctuations}

Let us consider a surface with an isolated step of monoatomic
height. The step position, at time $t$
and step site $x$, is specified by $h(x,t)$ (in units of the lattice
spacing, i.e., $h= ...-1, 0, 1,...$; likewise $x= 1,2,..., L$, where
$L$ is the length of the step). For
instance, $h(x,t)= 0$ describes a perfectly straight step. Step
fluctuations may be quantified by the correlation function
\begin{equation}
G(t)=<(h(x,t)- h(x,0))^2>_{x,ensemble}
\end{equation}
where the brackets denote spatial (along the step edge, $x$)
and ensemble averages. The initial step position, $h(x,0)$, may
be chosen in various ways, for instance, it may be straight or
thermally equilibrated. 

At fixed time, the step stiffness may be characterised by
the spatial correlation function
\begin{equation}
F(\Delta x)=<(h(x_0,t)- h(x_1,t))^2>_{ensemble}
\end{equation}
where $\Delta x$ is the distance between $x_0$ and $x_1$. Choosing
$x_0$ and $x_1$ in the
center and at the end of the step, respectively, $F$ then measures
the width of the step (and will be denoted by $w$). Usually, the
width of the step is
limited by and related to its length, $L$.

If each step position would be completely free to wander back
and forth with the same probability at any time (i.e. each
step position would be a random walker), then $G(t)$ would
obviously obey a diffusive behaviour, $G(t) \propto t$. This 
situation may be realized, e.g., at infinite temperature. However, in
general, the motion of each step position is hindered due to
the other parts of the step. Indeed, the step meandering is
typically subdiffusive, with a simple power law at late
times, $G(t) \propto t^a$, where the exponent $a < 1$. The
value of $a$ depends on the atomic process driving the 
step fluctuations, and $a$ may be determined theoretically or in
experiments.

In the theoretical approaches, assumptions are made on the
transition rates and correlations for the attachment and
detachment of the atoms at the step edges    
as well as on the boundary conditions
at the ends of the step (for example, periodic or pinned). In
particular, Langevin descriptions and Monte Carlo simulations
have been applied.

The basic equation of the Langevin theory for step dynamics has
the form (\cite{k:e1,bla:dux})
\begin{equation}
\partial h(x,t)/\partial t = {\cal J}(h(x,t)) +\eta(x,t)
\end{equation}
where the functional ${\cal J}$ describes the reduction of
the step free energy, and $\eta$ is the noise
term. Both quantities depend on the microscopic processes driving
the step fluctuations.
Usually, $h$ and $x$ are assumed to be continuous variables.
The Monte Carlo simulations (for an
introduction, see Binder (1992)) are performed for 
microscopic models, specifying
the interactions between the surface (or step) atoms and the transition rates
for the motion of the atoms at the surface, as shown in the following.

\subsection{Evaporation-Condensation}
Let us assume that the step fluctuations are due to uncorrelated
attachment and detachment of atoms at the
step edges (as may be realized when the surface dynamics is
governed by exchange of atoms with the surrounding vapour: kinetics
of evaporation and condensation, EC).

In a Monte Carlo simulation, the step may now be mimicked by a
one--dimensional SOS model
\begin{equation}
 {\cal H}= \sum\limits_{(x,x')} J\left|h(x)- h(x')\right|
\end{equation}
where the sum runs over all nearest neighbour pairs $(x,x')$
of step sites; $J$ is
the kink energy. A resulting step configuration is depicted in
Fig. 1.

\begin{figure}
\centerline{\epsfxsize=9.0cm \epsfbox{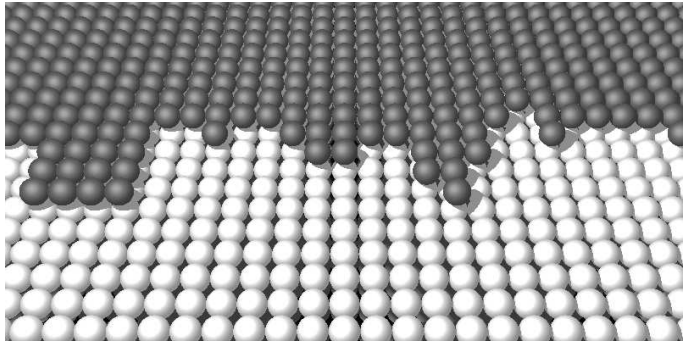}}
%\picplace{9.0cm}
\caption{Typical step configuration based on the one--dimensional
SOS model ($k_BT/J= 1.2$), using EC. Dark atoms constitute
the upper terrace, light ones the lower terrace, with the
step sites along the horizontal direction. Only part of the
step is shown}
\end{figure}

The EC kinetics corresponds, in the simulations, to Glauber
dynamics. The transition rate for attachment or
detachment of an atom at a randomly chosen step site is
given by the Boltzmann factor $e^{-\delta E/k_BT}$, with $\delta E$
being the energy change needed for the possible attachment or
detachment, $k_B$ is the Boltzmann constant, and $T$ the
temperature. Choosing the initial step positions $h(x,t=0)$, $G(t)$
is computed by averaging over a large number, $N$, of samples. The results
depend on the temperature, the number of active step sites, $L$, the boundary
conditions at the ends of the step, and the initial step
configuration.  
In particular, we monitored $G(t)$ and
the step width $w$ for steps of length $L$ ranging from
4 to 200 at various temperatures ($k_BT/J$ going from 0.5 to 4.0), pinning
the step positions
at the ends, $h(1,t)= h_1$ and $h(L+2,t)= h_2$ (usually, we
took $h_1= h_2= 0$), or using periodic boundary conditions (pbc) at the
ends of the steps. The time $t$ is measured in Monte Carlo
steps per site (MCS), i.e. one time unit has elapsed after $L$
attempts to change the step position by one. Averages have been taken over 
typically $N= 10^6$ to $10^7$ samples.

At very short times, the step fluctuations are
nearly diffusive, with $G(t) \propto t$, due to the random
events at (almost) equivalent step sites. However, caused by
the rigidity of the step, deviations show up
soon. They may be monitored conveniently by the effective
exponent
\begin{equation}
 a_{eff}(t)= d ln(G(t))/d ln(t)
\end{equation}
If $G(t)$ would obey a simple power law, the exponent would be a
constant, $a= a_{eff}$.

\begin{figure}
\centerline{\epsfxsize=8.0cm \epsfbox{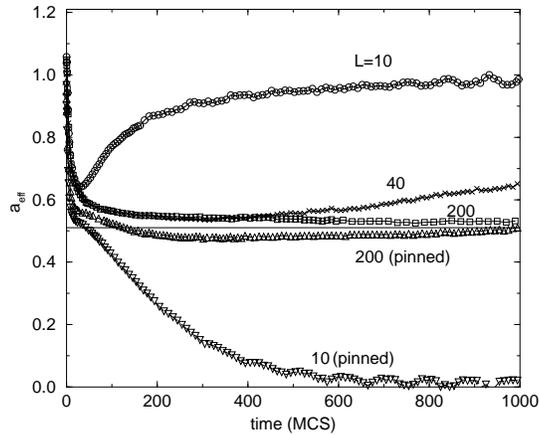}}
%\picplace{6.5cm}
\caption{Effective exponent of the step fluctuations, $a_{eff}$, versus
time, in MCS, for EC at $k_BT/J= 1.2$, using periodic boundary
conditions for initially straight steps (step length $L$= 10, 40, and
200), or with thermalized steps pinned at the
ends, $h_1= h_2= 0$, $L= 10$ and 200. The solid line refers to $a_{eff}=1/2$}
\end{figure}

As depicted in Fig. 2, $a_{eff}$ tends to decrease
with time (apart from
the behaviour of initially straight steps, where the exponent
first increases slightly to values larger than one before
getting smaller).
The rate of the decrease in $a_{eff}$
depends non--monotonically on temperature, being
fastest at some intermediate values, say, $k_BT \approx J$.
It is higher when using thermally equilibrated initial
step positions $h(x,t=0)$, compared to straight steps. For
sufficiently long steps the effective exponent seems to approach, at
large times, $a= 1/2$, the asymptotic value obtained
from Langevin descriptions for EC kinetics of indefinitely extended
steps (\cite{ed:wi,lipo:1,ab:up,sta:lan,b:e3,bla:dux}),
and observed experimentally, e.g., on a Au(110)
surface (\cite{ku:ho,ke:be}).

However, as illustrated in Fig. 2, the finite length of the step
leads to different asymptotics, depending on the boundary conditions. For
pbc, $a_{eff}$ may get close to 1/2, before
it rises at later times when the step
width $w$ saturates. The
time $t_{w}$, at which the step width has reached its saturation value
depends on the length of the step (exact
calculations for a simplified model
suggest $t_{w} \propto L^2$ (\cite{ab:up}), in reasonable agreement
with our simulations). Eventually, the step behaves effectively 
like a single random
walker with a reduced diffusion coefficient, i.e. 
$a_{eff} \longrightarrow 1$.

In the pinned case, the step width
also saturates, $w \propto L^{1/2}$ for
both boundary conditions (\cite{bla:dux}). Obviously, the
motion of the step is now hindered because the step 
is fixed at the ends. Therefore, $G(t)$ remains bounded and
$a_{eff}$ tends towards zero.

\subsection{Periphery Diffusion}
We now suppose that the step fluctuations are caused by direct hops of
step atoms between neighbouring step sites (in nature, the hops
correspond to the motion of individual atoms or they may result from a
particle exchange mechanism). The step or 
periphery diffusion (PD) may be
simulated again for a one--dimensional SOS model, equation
(4), applying now Kawasaki dynamics (\cite{bind:1}). The
transition rates for diffusion
of a randomly chosen step atom to a neighbouring site are
given by the Boltzmann factor of the energy change needed for the
hop. 

The Monte Carlo simulations for PD are done in complete analogy to
those for EC, studying the impact of the step length, the
initial step configuration, temperature, and the boundary conditions.
Most runs were done at $k_BT/J= 3.0$. Averages were taken over
$N= 10^6$ to $10^7$ samples. The step length $L$ ranged from 6 to 200.

At very short times, the step fluctuations are nearly diffusive, before
the rigidity of the step starts to slow down their growth.
The corresponding rate of decrease
in the effective exponent $a_{eff}$ of the correlation function $G(t)$ 
depends again in a non--monotonic
fashion on temperature. $a_{eff}$ seems
to approach the value 1/4 for indefinitely long steps, see
Fig. 3, in accordance with predictions from
Langevin descriptions (\cite{k:e1,bla:dux}) and experimental data for
steps on, e.g., Cu surfaces (\cite{gi:ib}).

However, the finite step length $L$ modifies that behaviour at late times, 
reflecting also the boundary conditions at the ends of the step, as
illustrated in Fig.3.
Pinning of the steps limits
the fluctuations, with $a_{eff}$ decreasing eventually
to zero when the width of the step has acquired its saturation
value. For periodic boundary conditions and short steps, the effective
exponent shows a rather peculiar time dependence, as depicted in
Fig. 3. The step fluctuations $G(t)$ are bounded, because the
average step position is, by definition of PD, conserved, in
contrast to the EC case where the entire
step of finite width may move like a random walker at late times.
Accordingly, $a_{eff}$ goes to zero. Note the pronounced
maximum in $a_{eff}$ for the short step ($L$ =10, at about 100 MCS, see
Fig. 3). This feature seems to be typical for longer steps as well (e.g., for
$L= 20$, the maximum shows up at about 1600 MCS). The faster growth in
the step fluctuations might be related to a modulation in the
step positions. Then the amplitudes of the 'hill' and
'valley' parts of the fluctuations may increase, satisfying the above
conservation requirement, until
finally these excursions get bounded
as well. This aspect, which we also noticed for pinned steps, deserves 
further studies. 

\begin{figure}
\centerline{\epsfxsize=8.0cm \epsfbox{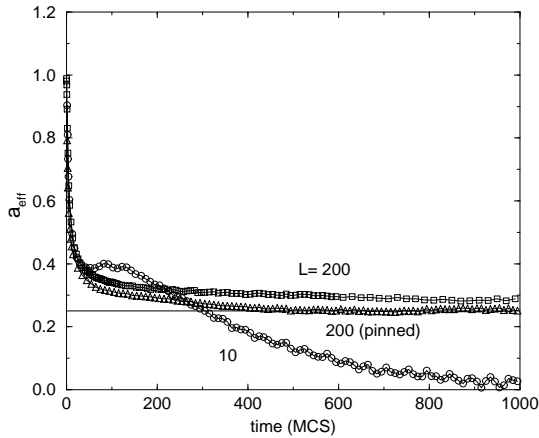}}
%\picplace{6.5cm}
\caption{Effective exponent of the step fluctuations, $a_{eff}$, versus
time, in MCS, for PD at $k_BT/J= 3.0$, using periodic boundary
conditions for initially straight steps (step length $L$= 10 and
200), or with thermalized steps pinned at the
ends, $h_1= h_2= 0$, $L= 200$. The solid line refers to $a_{eff}= 1/4$}
\end{figure}

In addition, we studied the situation where the step is pinned at some
angle, i.e., the ends of the steps are at different
positions, $h_1$, and $h_2$. The dynamics seems to be quite
similar to that of step where the pinning introduces no
additional kinks ($h_1= h_2$), at least at the fairly
high temperature which we considered, $k_BT/J= 3$. A more detailed
analysis seems to be desirable.

\subsection{Terrace Diffusion}
In general, the step dynamics on crystal surfaces is expected to be
rather complex, because atoms
detaching from a step may diffuse (anisotropically
and in external potentials) on the neighbouring lower and
upper terraces, colliding with other atoms on the terraces. In 
addition, exchange mechanisms may introduce
correlations of different ranges between atoms detaching from and
attaching at the steps. There have been a few attempts to 
study the complexity of these processes in the framework of
Langevin descriptions. A rather appealing picture
has been suggested by Duxbury and Blagojevic (1997), by expressing the
functional ${\cal J}$ of the Langevin equation in terms of the
chemical potential at the step sites and a 'diffusion kernel'
$P(l)$, which denotes the probability of an atom being originally
one lattice spacing in front of the step site $x$ to attach at
step position $x+l$, after having undergone an
appropriate walk on the terrace. The
approach has been critically discussed by Einstein
and Khare (1998). In any event, it poses the well defined
and interesting problem to calculate the diffusion kernel for
various scenarios.

For EC, P(l) would be obviously constant, i.e. the wandering atom could
attach with the same probability at each step site. Indeed, one
obtains, via equation (3), that the
step fluctuations obey, at late
times, the power--law $G(t) \propto t^{1/2}$. For PD, $P(l)$
vanishes for $l > 1$, yielding the asymptotic exponent
$a =1/4$. Perfect terrace diffusion (TD), where the atom executes
a standard random walk on the terrace in front of 
a straight step until it attaches at that reference step, is
described by a diffusion kernel, at
large distance $l$, of the form $P(l) \propto 1/l^2$. This 
decay law leads then to subdiffusive step fluctuations with
$G(t) \propto t^{1/3}$, being intermediate between EC and PD (\cite{bla:dux}).

The previous calculations of $P(l)$ for TD had been done in the continuum
limit for the step position $h(x,t)$. We extended
those calculations in several ways. $P(l)$ was
determined by an exact numerical enumeration of the probability
to visit a given, discrete site of the terrace (\cite{stan:1}), which
is supposed to be a square lattice. In doing that, we are dealing
with a lattice of finite extent, bordered by a step being parallel
and opposite (at a distance $d$) to the reference step and two steps 
starting at the ends and being 
perpendicular to the reference step of length $L$. The three boundary steps
may either
reflect or absorb the wandering adatom. The boundary conditions
allow one to study, for instance, in which way the step dynamics
changes because of the presence of a neighbouring
step with a large (reflecting) or small (absorbing) Schwoebel--Ehrlich
barrier.

In particular, we considered an external potential, V, of the form
\begin{equation}
 V= A/ y^2
\end{equation}
where $y$ measures the distance of the diffusing adatom from the reference
step; A (in units of $k_BT$) is the strength
of the potential originating from elastic or dipolar interactions of
the adatom with the reference step (and its opposite step). Thence, the
hopping probabitities for the wandering adatom are different for
moves towards and away from the steps. Assuming straight steps, the
resulting diffusion kernel $P(l)$ is shown to display the same limiting
behaviour as in the potential--free ($A= 0$) case, as illustrated
in Fig. 4. In particular, for large distances $d$ between the reference
and the opposite step, $P(l)$ falls off as $1/l^2$ for $l >> 1$. Only
the prefactor of the power--law
depends on the strength $A$ of the potential, increasing
exponentially with $A$ (at least for small values of $A$). Accordingly, the 
interaction of the adatom with the step may be expected to be irrelevant
for TD, with $G(t) \propto t^{1/3}$. For smaller distances $d$, $P(l)$ 
decreases exponentially with $l$. The exponential decay, due
to reflection or adsorption at 
the neighbouring step, leads to a slowing down of the
step fluctuations, being now in the limiting case of
PD, $G(t) \propto t^{1/4}$ (\cite{bla:dux}). Of course, it
would only hold if there is a reflecting (infinite) Schwoebel--Ehrlich
barrier, because otherwise the neighbouring (opposite) step would be a source
of adatoms which could attach at the reference step in an uncorrelated
manner, giving rise to the fast step fluctuations
of EC, $G(t) \propto t^{1/2}$.    

\begin{figure}
\centerline{\epsfxsize=8.0cm \epsfbox{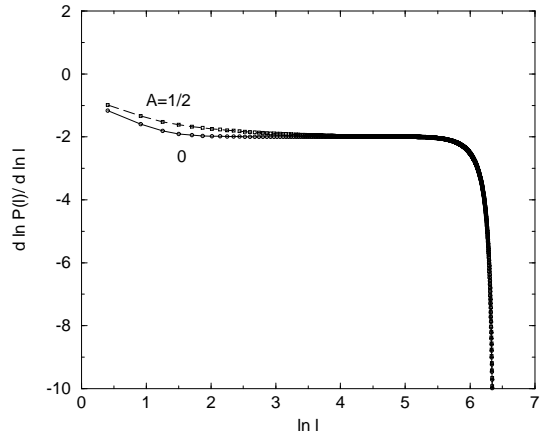}}
%\picplace{6.5cm}
\caption{Effective exponent of the diffusion
kernel, $d ln P(l)/d ln l$, versus return distance $l$, for
a wandering adatom without ($A= 0$) or
in an external potential, $V= A/y^2$. The opposite step is
reflecting, while the steps at the side are absorbing. The length
of the reference step is 1200, while the distance to the 
opposite step is 600. Note that the data are, on the scale of 
the figure, indistiguishable for large return distances. The plateau
corresponds to the $1/l^2$--law discussed in the text. The following
decay reflects the presence of the opposite step}
\end{figure}

Furthermore, we studied the situation, where the reference step is
rough. The asymptotics of $P(l)$ seems to be unaffected by
this modification as well (\cite{bisa:1}), e.g., if the 
influence of the opposite
step is sufficiently strong, we recover the same exponential decay
in $P(l)$ as for straight steps, with the prefactor reflecting the roughness.

The findings suggest that the corresponding 
time dependence of the step fluctuations at late
stages, $G(t) \propto t^{1/3}$, is rather
robust against including
elastic or dipolar interactions of the diffusing adatom with the step
and also against the roughness of the step. Different time laws, with
$a$ deviating from 1/3, may, however, result from a competition
between terrace and periphery diffusion with
various sticking coefficients (\cite{k:e1}) or collisions
between adatoms diffusing on the terrace. The topic may be studied
in simulations using, for instance, the kinetic Monte Carlo
approach with suitable activation energies (or using standard Monte Carlo
techniques for a two--dimensional SOS model with
a step (\cite{b:e3}), applying Kawasaki dynamics). Such simulations
have been done in particular for a model of steps of monoatomic
height on Ag(110) (\cite{ste:1}), motivated by recent
experiments (\cite{be:li}). The terrace diffusion is extremely
anisotropic. The steps may become quickly very fuzzy, and they
are no longer uniquely defined, so that rather involved
analyses are needed to identify the characteristic time--law
for the step fluctuations.
 
\section{Summary}
We considered three limiting cases for step fluctuations
at crystal surfaces, where the attachment and detachment
of atoms at the steps is (i) uncorrelated, EC, (ii) due to local
moves along the surface edge, PD, or (iii) mediated by
random walks of adatoms on the neighbouring terraces, TD.
In all three cases, the fluctuations of an indefinitely long step 
are expected to grow subdiffusively at large
times, $G(t) \propto t{^a}$, with $a$= 1/2, 1/4 and 1/3, respectively.

The first two cases have been studied by doing Monte Carlo simulations
for one--dimensional SOS models. We confirmed the generic 
time laws for step fluctuations at late stages, with
$a= 1/2$ for EC, and 1/4 for PD. However, due
to the finite length of the step and boundary effects at the
ends of the step, various other typical scenarios are 
encountered, including the crossover from subdiffusive to
diffusive behaviour, for EC and periodic boundary conditions.

The case of terrace diffusion has been discussed in the framework
of a recent Langevin description. The main quantity is the
diffusion kernel, describing the return
probability of an adatom wandering on the terrace to a step site.
Its asymptotics, which is supposed to determine the exponent
$a$ (being 1/3 for perfect terrace diffusion), is found to be
robust against realistic interactions of the adatom with
the step as well as against roughening of the step. 

Situations in which deviations from these limiting cases may be
possible have been mentioned.

%
% ---- Bibliography ----
%


\begin{thebibliography}
%
\bibitem{}{ab:up}{Abraham and Upton (1989)}
Abraham,\,D. B., Upton,\,P. J. (1989):
Dynamics of Gaussian interface models.
Phys.\, Rev.\, B {\bf 39}, 736
%
\bibitem{}{b:e3}{Bartelt et al. (1994)}
Bartelt,\,N. C., Einstein,\,T. L., Williams, E.D.  (1994):
Measuring surface mass diffusion coefficients by observing step fluctuations.
Surf.\, Sci.\, {\bf 312}, 411
%
\bibitem{}{bind:1}{Binder (1992)}
Binder,\,K., (ed.) (1992):
The Monte Carlo method in Condensed Matter Physics.
(Springer, Berlin,Heidelberg)
%
\bibitem{}{bisa:1}{Bisani (1998)}
Bisani,\,M. (1998):
Zur Theorie wechselwirkender Stufen auf Kristalloberfl\"achen.
Diplomarbeit, RWTH Aachen
%
\bibitem{}{bla:dux}{Blagojevic and Duxbury (1997)}
Blagojevic,\,W., Duxbury,\,P. M. (1997):
From atomic diffusion to step dynamics. In:
Dynamics of crystal surfaces and interfaces. Eds. Duxbury,\,P. M.,
Pence,\,T. J. (Plenum Press, New York, London), p.1
%
\bibitem{}{bon:zel}{Bonzel and Surnev (1997)}
Bonzel,\,H. P., Surnev,\,S. (1997):
Morphologies of periodic surface profiles and small particles. In:.
Dynamics of crystal surfaces and interfaces. Eds. Duxbury,\,P. M.,
Pence,\,T. J. (Plenum Press, New York, London), p.41
%
\bibitem{}{ed:wi}{Edwards and Wilkinson (1982)}
Edwards,\,S. F., Wilkinson,\,D. R. (1982):
The surface statistics of a granular aggregate.
Proc.\, R.\, Soc.\, A {\bf 381}, 17
%
\bibitem{}{gi:ib}{Giesen--Seibert et al. (1993)}
Giesen--Seibert,\,M., Jentjens,\, R., Poensgen,\,M., Ibach,\,H. (1993):
Time dependence of step fluctuations on vicinal Cu ( 1 1 19) surfaces
investigated by tunneling microscopy.
Phys.\, Rev.\, Lett.\. {\bf 71}, 3521
%
\bibitem{}{ha:sp}{Hager and Spohn (1995)}
Hager,\,J., Spohn,\,H. (1995):
Self--similar morphology and dynamics of periodic surface profiles
below the roughening transition.
Surf.\, Sci.\, {\bf 324}, 365
%
\bibitem{}{k:e2}{Khare et al. (1996)}
Khare,\,S. V., Bartelt,\,N. C., Einstein,\,T. L. (1996):
Brownian motion and shape fluctuations of single--layer adatom and
vacancy clusters on surfaces: Theory and simulations.
Phys.\, Rev.\, B {\bf 54}, 11752
%
\bibitem{}{k:e1}{Khare and Einstein (1998)}
Khare,\,S. V., Einstein,\,T. L. (1998):
Unified view of step-edge kinetics and fluctuations.
Phys.\, Rev.\, B {\bf 57}, 4782
%
\bibitem{}{ku:ho}{Kuipers et al. (1993)}
Kuipers,\,L., Hoogeman,\,M. S., Frenken,\,J. W. M. (1993):
Step dynamics on Au(110) studied with a high--temperature, high--speed
scanning tunneling microscope.
Phys.\, Rev.\, Lett.\. {\bf 71}, 3517
%
\bibitem{}{vil:1}{Lancon and Villain (1990)}
Lancon,\,F., Villain,\,J. (1990):
Dynamics of a crystal surface below its roughening. In:
Kinetics of ordering and growth at surfaces. Ed. Lagally,\,M. G.
(Plenum Press, New York, London), p.369
%
\bibitem{}{be:li}{Li et al. (1996)}
Li,\,J., Berndt,\,R., Schneider,W.--D., (1996):
Tip--assisted diffusion on Ag(110) in scanning tunneling microscopy.
Phys.\, Rev.\, Lett.\. {\bf 76}, 11
%
\bibitem{}{lipo:1}{Lipowsky (1985)}
Lipowsky,\,R. (1985):
Nonlinear growth of wetting layers.
J.\, Phys.\, A {\bf 18}, L 585
%
\bibitem{}{stan:1}{Majid et al. (1984)}
Majid,\,I., Ben--Avraham,\, D., Havlin,\,S., Stanley,\,H. E. (1984):
Exact enumeration approach to random walks on percolation clusters
in two dimensions.
Phys.\, Rev.\, B {\bf 30}, 1626
%
\bibitem{}{mul:1}{Mullins (1959)}
Mullins,\,W. W. (1959):
Flattening of nearly plane solid surface due to capillarity.
J.\, Appl.\, Phys.\, {\bf 30}, 77
%
\bibitem{}{pi:1}{Pimpinelli et al. (1993)}
Pimpinelli,\,A., Villain,\,J., Wolf,\, D. E., Metois,\,J. J., Heyraud,\,J.
 C., Elkiani,\,I., Uimin,\,G. (1993):
Equilibrium step dynamics on vicinal surfaces.
Surf.\, Sci.\, {\bf 295}, 143
%
\bibitem{}{sel:1}{Selke and Duxbury (1995)}
Selke,\,W., Duxbury,\,P. M. (1995):
Equilibration of crystal surfaces.
Phys.\, Rev.\, B {\bf 57}, 4782
%
\bibitem{}{sta:lan}{Stauffer and Landau (1989)}
Stauffer,\,D., Landau,\,D. P. (1989):
Interface growth in a two--dimensional Ising model.
Phys.\, Rev.\, B {\bf 57}, 4782
%
\bibitem{}{ste:1}{Stebens (1998)}
Stebens,\,A. (1998):
Monte--Carlo Simulationen von dynamischen Prozessen
auf Kristalloberfl\"achen.
Diplomarbeit, RWTH Aachen
%
\bibitem{}{ke:be}{van Beijeren et al. (1983)}
van Beijeren,\,H., Kehr,\,K. W., Kutner,\,R. (1983):
Diffusion in concentrated lattice gases. III. Tracer diffusion on
a one--diemnsional lattice.
Phys.\, Rev.\, B {\bf 28}, 5711
%
\bibitem{}{w:1}{Weeks et al. (1997)}
Weeks,\,J. D., Liu,\,D.--J., Jeong,\, H.--C. (1997):
Two--dimensional models for step dynamics. In:.
Dynamics of crystal surfaces and interfaces. Eds. Duxbury,\,P. M.,
Pence,\,T. J. (Plenum Press, New York, London), p.41
%
\bibitem{}{wi:1}{Williams (1994)}
Williams,\,E. D. (1994):
Surface steps and surface morphology: understanding macroscopic
phenomena from atomic observations.
Surf.\, Sci.\, {\bf 299/300}, 502
%
\end{thebibliography}
\end{document}